\documentclass[aps,prl,twocolumn]{revtex4}
\usepackage{amssymb}
\usepackage{graphicx,amsmath}

\newcommand{\bea}{\begin{eqnarray}}
\newcommand{\eea}{\end{eqnarray}}
\newcommand{\beq}{\begin{equation}}
\newcommand{\eeq}{\end{equation}}
\newcommand{\benu}{\begin{enumerate}}
\newcommand{\enu}{\end{enumerate}}

\newcommand{\al}{\alpha}

\newcommand{\om}{\omega}

\newcommand{\ep}{\epsilon}

\newcommand{\si}{\sigma}

\newcommand{\ham}{\mathcal{H}}

\newcommand{\ptl}{\partial}

\newcommand{\bk}{{\bf k}}
\newcommand{\bq}{{\bf q}}

\begin{document}

\title{Variation of shear moduli across superconducting phase transitions}
\date{\today}
\author{Dimitri Labat,$^1$ Panagiotis Kotetes,$^{2,3}$ Brian M. Andersen,$^2$ and Indranil Paul$^1$}
\affiliation{
$^1$Laboratoire Mat\'{e}riaux et Ph\'{e}nom\`enes Quantiques, Universit\'{e} de Paris, CNRS,
F-75013, Paris, France\\
$^2$Niels Bohr Institute, University of Copenhagen, DK-2100 Copenhagen, Denmark\\
$^3$CAS Key Laboratory of Theoretical Physics, Institute of Theoretical Physics,
Chinese Academy of Sciences, Beijing 100190, China
}

\begin{abstract}
We study how shear moduli of a correlated metal change across superconducting phase transitions.
Using a microscopic theory we explain why for most classes of superconductors this change is small.
The Fe-based and the A15 systems are notable exceptions where the change is boosted by five orders
of magnitude. We show that this boost is a consequence of enhanced nematic correlation. The theory
explains the unusual temperature dependence of the orthorhombic shear and the back-bending of the
nematic transition line in the superconducting phase of the Fe-based systems.
\end{abstract}


\maketitle
An important topic in the field of high temperature
superconductivity is to understand the interplay between superconducting and nematic orders.
The issue arises naturally for the Fe-based
systems whose phase diagram shows a ubiquitous presence of the two
orders~\cite{review-feas,kuo-fisher-2016,chowdhury-2011,Livanas,
fernandes-2013,glasbrenner-2015,mishra-2016,gallais-2016,sprau-2017,andersen-2017,classen-2017,benfatto-2018}.
The relevance of nematicity to understand the pseudogap state of the cuprate superconductors
is currently under active investigation as
well~\cite{ando-2002,howald-2003,hinkov-2008,daou-2010,sato-2017,auvray-gallais-2019,kivelson-2003,vojta-2009,wang-2013}.

One cause of interplay is fluctuations associated with the two orders, and the effect of
nematic fluctuations on superconductivity has been extensively
studied in the past~\cite{yamase-2013,maier-2014,metlitski-2015,lederer-2015,labat-2017,klein-2018,lederer-2019}.
A second cause can be a third degree of freedom such as antiferromagnetic fluctuations which can enhance
nematic correlation, but which are themselves suppressed in a singlet superconductor~\cite{fernandes-2010}.
What is less examined is the effect of the superconducting order itself on the nematic properties of electrons in
solids. The goal of the current paper is to study the last from a microscopic point of view.

For such a study a shear strain of a suitable symmetry is an appropriate nematic order parameter, even if the
nematic transition is driven by electronic interactions~\cite{fernandes-2014,gallais-review-2016}.
This is because, due to electron-strain coupling,
the nematic transition at temperature $T_s$ itself manifests as a structural instability. Consequently, tracking the
shear elastic constant $c_s(T)$ as a function of temperature $T$, especially across the superconducting transition at
$T_c$, is a practical method to study the interplay. For simplicity we restrict to the case
where $T_c > T_s$.

More concretely, for $T \sim T_c$, the free energy per unit volume
involving the shear strain $u_s$ and the superconducting order parameter $\Delta$  can be written as
\beq
\label{eq:F}
F = (c_s/2) u_s^2 + (a/2) \left|\Delta\right|^2 +  (b/4) \left|\Delta\right|^4 + (\lambda/2) u_s^2 \left|\Delta\right|^2.
\eeq
Here $\Delta$ has dimension of energy, while $(a, \lambda)$ have that of density of states (DOS),
$a = a_0 (T - T_c)$, and $b>0$.
The fourth term, which captures the interplay, describes
how the shear elastic constant is modified across $T_c$.
In the above we assumed that $\Delta$ belongs to a one dimensional irreducible representation of the
unit cell point group, and that
there is no second nearly critical symmetry channel for superconductivity~\cite{Livanas,fernandes-2013,kushnirenko-2018}.

From Eq.~(\ref{eq:F}) it follows that $c_s(T)$ itself is
continuous at $T_c$, but its temperature derivative jumps at $T_c$ with the jump given by
$(d c_s/dT)_{T_c^+} - (d c_s/dT)_{T_c^-} = \lambda a_0/b$.
In other words, $c_s(T)$ has a kink at $T_c$ which encodes information about the interplay parameter $\lambda$.
The magnitude of this kink can be quantified by
$\delta c_s/|c_s^m|$, where $\delta c_s \equiv \lambda \Delta_0^2 \sim c_s^{s} - c_s^m$. Here $c_s^{s}$ is the zero
temperature elastic constant in the superconducting phase, $c_s^m$ is
inferred from the $T=0$ extrapolation of $c_s(T)$ in the metal phase, and $\Delta_0 \equiv \Delta(T=0)$.

A literature search reveals that in most known classes of superconductors the
ratio $\delta c_s/|c_s^m|$ is ``small'' and is of order $10^{-6}$.
Examples include conventional Bardeen-Cooper-Schrieffer (BCS) systems~\cite{olsen,alers61},
cuprates such as La$_{2-x}$Sr$_x$CuO$_4$ at various dopings
(see Figs.\ 7 and 8 in Ref.~\cite{nohara}), and heavy fermion systems UPt$_3$ and URu$_2$Si$_2$~\cite{bruls,thalmeier}.
From Ehrenfest-type thermodynamic argument it is known that  $\delta c_s/|c_s^m|$
is related to the ratio between the superconducting condensation energy and the Fermi energy,
which is typically small~\cite{testardi75,millis-rabe}.
This provides a simple way to understand this small ratio without a microscopic analysis.

However, there are two classes of superconductors, namely the Fe-based~\cite{fernandes-2010,yoshizawa,zvyagina13,boehmer-2014,boehmer-2016}
and the A15 systems~\cite{testardi-67,rehwald-72,testardi-rmp},
for which this ratio is ``large'' with $\delta c_s/|c_s^m| \sim 10^{-1}$.
Clearly, this increase of $\delta c_s/|c_s^m|$ by \emph{five} orders of magnitude compared to the standard behavior
cannot be understood purely from thermodynamics,
and a microscopic approach is needed.
With this motivation, here we develop such a microscopic theory of the coupling $\lambda$ that encodes the interplay
between the two orders.

Our main results are the following. (i) First, we show that in systems with negligible nematic correlation
$\lambda/\mathcal{N}_F$ is small, where $\mathcal{N}_F$ is DOS at Fermi level. This is
due to a cancellation of the low-energy electronic contribution that is not imposed by symmetry.
We show that this cancellation is related to the general property that the quadrupolar charge susceptibility of an electronic system
remains approximately unchanged between its metallic and superconducting phases. This explains the small ratio of
$\delta c_s/|c_s^m|$ for most superconductors. (ii) Second, we show that for systems with large nematic correlation length
$\xi \gg l$, where $l$ is the interatomic distance, the parameter
$\lambda$ is boosted by $(\xi/l)^4$.
This accounts for the five orders of magnitude increase in $\delta c_s/|c_s^m|$ seen in
the A15 and the Fe-based systems.
(iii) Third, we show that the sign of $\lambda$, that controls cooperation or competition between the two orders, is non-universal and that it
depends on the band structure.
\begin{figure}[!!t]
\begin{center}
\includegraphics[width=8cm,trim=10 0 0 0]{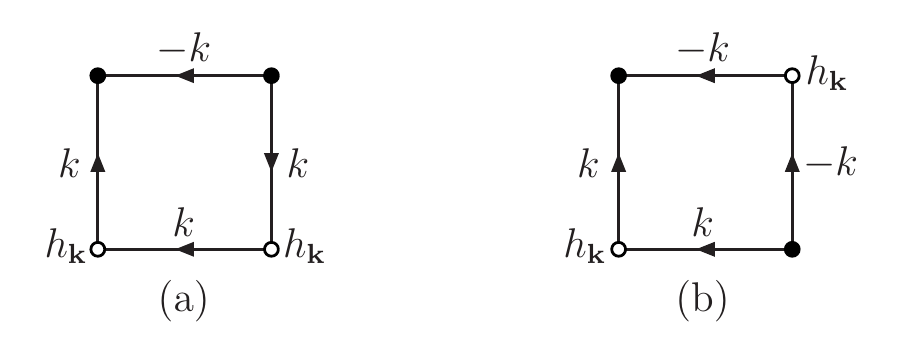}
\caption{
Diagrammatic representation of the coupling $\lambda$ that controls the interplay between superconducting
and nematic orders, see Eq.~(\ref{eq:F}). $\lambda$ is a four-point function with
two particle-hole (open circles) vertices with nematic form factor $h_{\bf k}$
and two particle-particle (closed circles) vertices, see  Eq.~(\ref{eq:lambda-1}).
Solid lines are electron Green's functions. $k = ({\bf k}, \omega)$ denote momentum and frequency.
}
\label{fig1}
\end{center}
\end{figure}

\emph{Microscopic theory.}
Our main message can be illustrated by considering a one-band metal in a tetragonal lattice.
The relevant elastic constant can be written as
\beq
\label{eq:elastic-constant}
c_s(T) \equiv c_0 - \al^2 \chi_n (T).
\eeq
$c_0$ is the modulus of the bare elastic medium, which we assume to be temperature independent.
$\alpha$ is the electron-strain interaction energy, such that in the presence of a finite strain the electron dispersion
changes as $\epsilon_{\bk} \rightarrow \tilde{\epsilon}_{\bk} = \epsilon_{\bk} + \alpha u_s h_{\bk}$.
To be concrete we take $u_s$ to be the orthorhombic strain that transform as $(x^2-y^2)$, in which case
$h_{\bk} \sim \cos k_x - \cos k_y$. The precise nature of the shear mode and the
associated form factor is unimportant.
Likewise, the spatial symmetry of $\Delta$ (i.e., $s$-, $p$- or $d$-wave) play no
role, and we take it as $s$-wave for simplicity.
The quantity $\chi_n \equiv \lim_{\bq \rightarrow 0} \chi_n(\bq, \omega =0)$,
where $\chi_n(\bq, 0)$ is the static nematic susceptibility of the electrons.
Thus, the role of the lattice variables is simply to probe the electronic properties, in particular
how $\chi_n$ changes across $T_c$.

At this point it is convenient to distinguish the following two situations.

\emph{(a) Away from nematic instability.}
When the system is far away from nematic/orthorhombic instability
the nematic correlation length is negligible, and therefore
$\chi_n^{s/m}(\bq, 0) \approx  \Pi_n^{s/m}(\bq, 0)$,
where $\Pi_n^{s/m}(\bq, 0)$ is the bare nematic susceptibility. We added superscripts $(s,m)$ to denote superconducting
and metallic phases, respectively.
In the superconducting phase the bare
nematic susceptibility is
\begin{align}
\Pi_n^{s}(\bq,0) &= - \frac{2}{\beta V} \sum_{\om_n, \bk} f^2_{\bk,\bq}
[G_{\bk + \bq} (i \om_n) G_{\bk} (i \om_n) \nonumber \\
&- F_{\bk + \bq} (i \om_n) F_{\bk} (i \om_n)],
\nonumber
\end{align}
where $\beta$ is inverse temperature, $V$ is volume, $f_{\bk,\bq} \equiv (h_{\bk} + h_{\bk + \bq})/2$
is the nematic form factor,
$G_{\bk}(i \om_n) = - (i \om_n + \ep_{\bk})/(\om_n^2 + E_{\bk}^2)$, $F_{\bk}(i \om_n) = \Delta/(\om_n^2 + E_{\bk}^2)$, and
$E_{\bk} = \sqrt{\ep_{\bk}^2 + \Delta^2}$. An overall factor two is due to spins. The equivalent expression for $\Pi_n^{m}(\bq,0)$
is obtained by setting $\Delta = 0$.

Eqs.~(\ref{eq:F}) and (\ref{eq:elastic-constant}) give
$
\lambda = - (\al^2/2) [\ptl^2 \chi_n^s/(\ptl |\Delta|^2)]_{|\Delta|=0}$.
Thus, $\lambda$ is a four-point function that can be obtained from $\Pi_n^m(0,0)$ by inserting two
particle-particle vertices (see Fig.~\ref{fig1}). This leads to the microscopic expression
\begin{align}
\label{eq:lambda-1}
\lambda = \lambda_0 &\equiv - \frac{2 \alpha^2}{\beta V}   \sum_{\om_n, \bk} h^2_{\bk}
\left[ 2 G^0_{\bk} (i \om_n)^3 G^0_{\bk} (- i \om_n) \right. \nonumber \\
&+ \left. G^0_{\bk} (i \om_n)^2 G^0_{\bk} (- i \om_n)^2 \right],
\end{align}
with
$
G^0_{\bk} (i \om_n)^{-1} \equiv i \om_n - \epsilon_{\bk}.
$
The above frequency sum is simple to perform. We define
the $B_{1g}$ density of states as
$
\mathcal{N}_{B_{1g}}(\epsilon) \equiv (1/V) \sum_{\bk} h^2_{\bk} \, \delta (\epsilon - \epsilon_{\bk}),
$
and we get
\[
\lambda_0 = - \frac{\alpha^2}{2} \int_{- \infty}^{\infty} d \epsilon
\mathcal{N}_{B_{1g}}(\epsilon) \frac{d^2}{d \epsilon^2}
\left[ \tanh (\beta \epsilon/2)/\epsilon \right].
\]

We expand $\mathcal{N}_{B_{1g}}$ around the Fermi energy as
$\mathcal{N}_{B_{1g}}(\epsilon) \approx \mathcal{N}_{B_{1g}}(0) + \epsilon \mathcal{N}_{B_{1g}}^{\prime}(0)
+ (\epsilon^2/2) \mathcal{N}_{B_{1g}}^{\prime \prime}(0) + \cdots$, where primes imply derivatives with respect
to energy.
Remarkably, the term proportional to $\mathcal{N}_{B_{1g}}(0)$, which is the contribution from the
low-energy excitations, \emph{vanishes}.
Since the term proportional to $\mathcal{N}_{B_{1g}}^{\prime}(0)$ is trivially zero, the first non-zero
contribution is
proportional to  $\mathcal{N}_{B_{1g}}^{\prime \prime}(0)$. We get,
\beq
\label{eq:lambda-4}
\lambda_0 = - \alpha^2 \mathcal{N}_{B_{1g}}^{\prime \prime}(0) \left\{ \log [\Lambda/(2T)] + \mathcal{C}_1 \right\},
\eeq
where $\mathcal{C}_1 \equiv \gamma - 3/2 - \log(\pi/4)$, $\gamma$ is the Euler constant, and $\Lambda$ is a high-temperature
cutoff. The logarithmic temperature
dependence above has the same origin as the familiar $\log (T)$ dependence of the particle-particle susceptibility in BCS theory.

The cancellation of the low-energy electronic contribution is important, and consequently it is useful to understand better
its physical origin.
Clearly, the cancellation is not dictated by any symmetry. Instead, it is a consequence of the
property that the \emph{bare} quadrupolar charge susceptibility of electrons
remains nearly unchanged across a metal to superconductor transition.
This can be demonstrated by the following calculation.

The frequency sum in the expression for $\Pi_n^{s}(0,0)$ gives
\[
\Pi_n^{s}(0,0) = \frac{1}{V} \sum_{\bk} h_{\bk}^2 \frac{\ptl}{\ptl \ep_{\bk}}
\left[ \frac{\ep_{\bk}}{E_{\bk}} \tanh \frac{E_{\bk}}{2T} \right].
\]
If we neglect the energy dependence of the $B_{1g}$ density of states $\mathcal{N}_{B_{1g}}(\epsilon)$, which is
appropriate for the low-energy electronic contribution, after the energy integral we get
\beq
\label{eq:approx-equality}
\Pi_n^{s}(0,0)_{\rm low} = \Pi_n^m(0,0)_{\rm low} = 2 \mathcal{N}_{B_{1g}}(0).
\eeq
In the above the subscript ``low'' implies the low-energy contribution.
In other words, from the perspective of the low-energy electrons
$\Pi_n^{s}(0,0)$ is independent of $\Delta$.
This property is
reminiscent of that of the uniform charge susceptibility $\ptl n/\ptl \mu$, where $n$ is the electron density and
$\mu$ the chemical potential. It is known that the Thomas-Fermi screening length, which is controlled by the
uniform charge susceptibility, remains practically unchanged when a metal turns into a superconductor~\cite{koyama04}.
The above discussion implies that if $\Pi_n^{s}(0,0)$ is expanded around $\Pi_n^m(0,0)$ in powers of $|\Delta|^2$, order by order the
pre-factors would be zero if we neglect the energy dependence of  $\mathcal{N}_{B_{1g}}(\epsilon)$. The coupling $\lambda_0$
in Eq.~(\ref{eq:lambda-1}) is related to the prefactor at order $|\Delta|^2$ in this expansion.

The above low-energy cancellation has the following consequences.
(i) Most importantly, we conclude that for superconductors with negligible nematic correlation
$\delta c_s/|c_s^m| \sim (T_c/E_F)^2$, where $E_F$ is the Fermi energy. This follows from the estimate
$\mathcal{N}_{B_{1g}}^{\prime \prime}(0) \sim \mathcal{N}_F/E_F^2$,
and by estimating the electron-phonon interaction energy
$\alpha$ as the geometric mean of the typical electronic and elastic energy scales, i.e., $(\alpha^2 \mathcal{N}_F/c_s) \sim 1$~\cite{agd}.
Thus, the above estimation, backed by a microscopic calculation, explains the order of magnitude of $\delta c_s/|c_s^m|$
reported for most known superconductors, the Fe-based and the A15 systems being exceptions.
(ii) The sign of $\lambda_0$, which governs whether the two orders cooperate or compete,
is non-universal and it depends on the sign of $\mathcal{N}_{B_{1g}}^{\prime \prime}(0)$.
(iii) Due to the absence of the low-energy contribution the coupling $\lambda \sim \lambda_0$ is nearly temperature independent.
This is consistent with the weak $T$-dependence of $\chi_n$ of several Fe-based systems at doping
well away from the nematic instability reported from elastoresistivity~\cite{chu2012} and
electronic Raman studies~\cite{gallais2013,blumberg2016}.

\emph{(b) Near a nematic instability.}
The above considerations need modification if the system is in the vicinity of a nematic instability, and the nematic
correlation length $\xi \gg l$, where $l$ is the interatomic distance.
For the sake of simplicity we assume that the nematic instability is a Pomeranchuk transition, i.e.,
spontaneous deformation of the Fermi surface.
Accordingly, we postulate the presence of an interaction
$
\ham_I = - (g/2) \sum_{\bq} O_n(-\bq) O_n (\bq),
$
with $g > 0$ having dimension of inverse DOS, and where
$
O_n(\bq) \equiv \frac{1}{\sqrt{V}} \sum_{\bk, \si} f_{\bk, \bq} c^{\dagger}_{\bk + \bq, \si} c_{\bk, \si},
$
is the quadrupolar charge operator. Such a phenomenological interaction has been widely used to study nematic instability
in metals~\cite{yamase-2013,lederer-2015,gallais-review-2016,gallais-2016,klein-2018,andersen-2019}.
In this case the increase of the nematic correlation length $\xi(T)$ with lowering temperature
can be described using random phase approximation, and the nematic susceptibility can be written as
$
\chi_n^i (\bq, 0) = \Pi_n^i(\bq, 0)/\left[ 1 - g \Pi_n^i(\bq, 0) \right],
$
where $i= (s,m)$.
As in case (a) we have
$
\lambda \propto [\ptl^2 \chi_n^s/(\ptl |\Delta|^2)]_{|\Delta|=0}$, and taking into account that
$\ptl \Pi_n^s(0,0)/\ptl |\Delta| =0$ due to gauge invariance, we conclude
\beq
\label{eq:lambda-renorm}
\lambda = \lambda_{\rm renorm} \equiv \lambda_0 (\xi/l)^4,
\eeq
where $(\xi/l)^2 = 1/[1 - g \Pi_n^m(0,0)]$.
From the above Eq.\ we deduce the following.
(i) Close to a nematic instability $g \Pi_n^m \sim 1$, or equivalently $\xi \gg l$, and therefore $\lambda$ and eventually
$\delta c_s/|c_s^m|$ can be boosted by
five orders of magnitude, even though the bare coupling $\lambda_0$ is small.
Note, the identification that electronic nematic correlation is significant in the A15 systems is an important conclusion
of our study.
(ii) In the metal phase the nematic
susceptibility $\chi_n^m(T) \propto (\xi(T)/l)^2 \sim \mathcal{N}_F \Lambda/(T - T_0)$. Here
$T_0$ is the nematic transition temperature of the electron-only subsystem,
with $T_0 = T_s - \al^2 \Lambda \mathcal{N}_F/c_0 \lesssim T_s$.
This implies that the renormalized $\lambda$ has power-law temperature dependence with
$\lambda_{\rm renorm}  \propto (\xi(T)/l)^4 \propto 1/(T - T_0)^2$. This is to be contrasted with case (a) where the bare
coupling $\lambda_0$ has weak logarithmic $T$-dependence.
\begin{figure}[!!t]
\begin{center}
\includegraphics[width=8cm,trim=10 0 0 0]{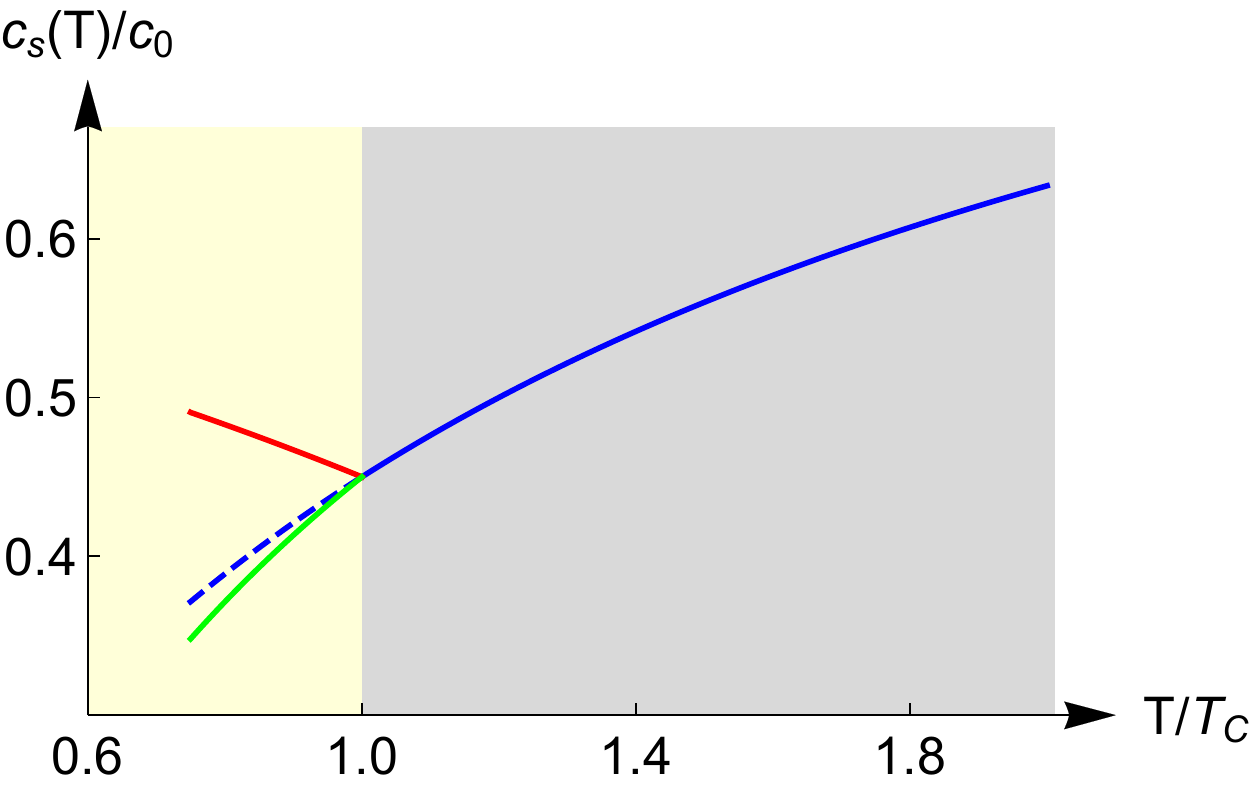}
\caption{
Kink in the $T$-dependence of the shear elastic constant $c_s(T)$ at a superconducting transition $T_c$.
The system is close to a nematic instability and the nematic correlation length
increases with lowering $T$. For sufficiently large $\lambda_0 >0$ the elastic constant hardens immediately
upon entering the superconducting phase (red/dark line), as seen in the Fe-based systems. For $\lambda_0 < 0$
the elastic constant softens more rapidly in the superconducting phase (green/light line). The dashed line is
the extrapolation of the metallic behavior.
}
\label{fig2}
\end{center}
\end{figure}

The enhancement of $\lambda$ implied by Eq.~(\ref{eq:lambda-renorm}) has the following two consequences.

\emph{1. $c_s(T)$ across superconducting $T_c$}.
Since $\chi_n^m(T) \propto 1/(T - T_0)$ while $\lambda_{\rm renorm} \propto 1/(T - T_0)^2$ has a stronger $T$-dependence,
it is clear that,
for $\lambda_0$ above a positive threshold,
the softening of $c_s(T)$ in the metal phase will turn into a hardening in the superconducting phase.
This can be illustrated from the following phenomenological modeling. We write
$
c_s(T)/c_0 = 1 - a_0 P(T)/[1- b_0 P(T)],
$
where $a_0 \equiv \al^2 \mathcal{N}_F/c_0$ and $b_0 \equiv g \mathcal{N}_F$
are constants, and $P(T)$ is the dimensionless bare nematic polarization. In the metallic phase we postulate
$P(T \geq T_c) = \Lambda/(T + T_1)$, with $T_1 \gg T_c$ such that $P(T)$ is weakly $T$-dependent around $T_c$.
As noted above, in the superconducting phase the bare polarization has an additional term proportional to
$\lambda_0 \Delta(T)^2$. We assume the mean field scaling $\Delta(T)^2 = \Delta_0^2(1-T/T_c)$,
and we write the bare interplay coupling $\lambda_0$
in terms of a dimensionless parameter $t_2 \equiv \lambda_0 \Delta_0^2/(\al^2 \mathcal{N}_F)$. This implies
$P(T \leq T_c) = \Lambda/(T + T_1) - t_2(1-T/T_c)$. It follows that, for sufficiently large
and positive $t_2 > \Lambda T_c/(T_c + T_1)^2$, the elastic constant $c_s(T)$ starts hardening
immediately below $T_c$, as seen in electron and holed doped
BaFe$_2$As$_2$~\cite{fernandes-2010,yoshizawa,boehmer-2014,boehmer-2016}. On the other hand, for
$t_2 <0$ (or equivalently $\lambda_0 <0$) the elastic softening enhances in the superconducting phase.
It is likely that this latter trend is relevant for FeSe$_{1-x}$S$_x$ at large doping where $T_c > T_s$~\cite{wang-2016}.
These two trends
are illustrated in Fig.~(\ref{fig2}) for which we use
$a_0 = 0.22$, $b_0 = 49.8$, $T_1/\Lambda=50$, $T_c/\Lambda =0.2$, while $t_2 = 1.3 \times 10^{-4}$ and
$t_2 = - 0.2 \times 10^{-4}$ for the red (dark) and green (light) lines, respectively.
For intermediate values of $t_2$ the $T$-dependence of $c_s(T)$ interpolates between these two limiting behaviors.

\begin{figure}[!!t]
\begin{center}
\includegraphics[width=8cm,trim=10 0 0 0]{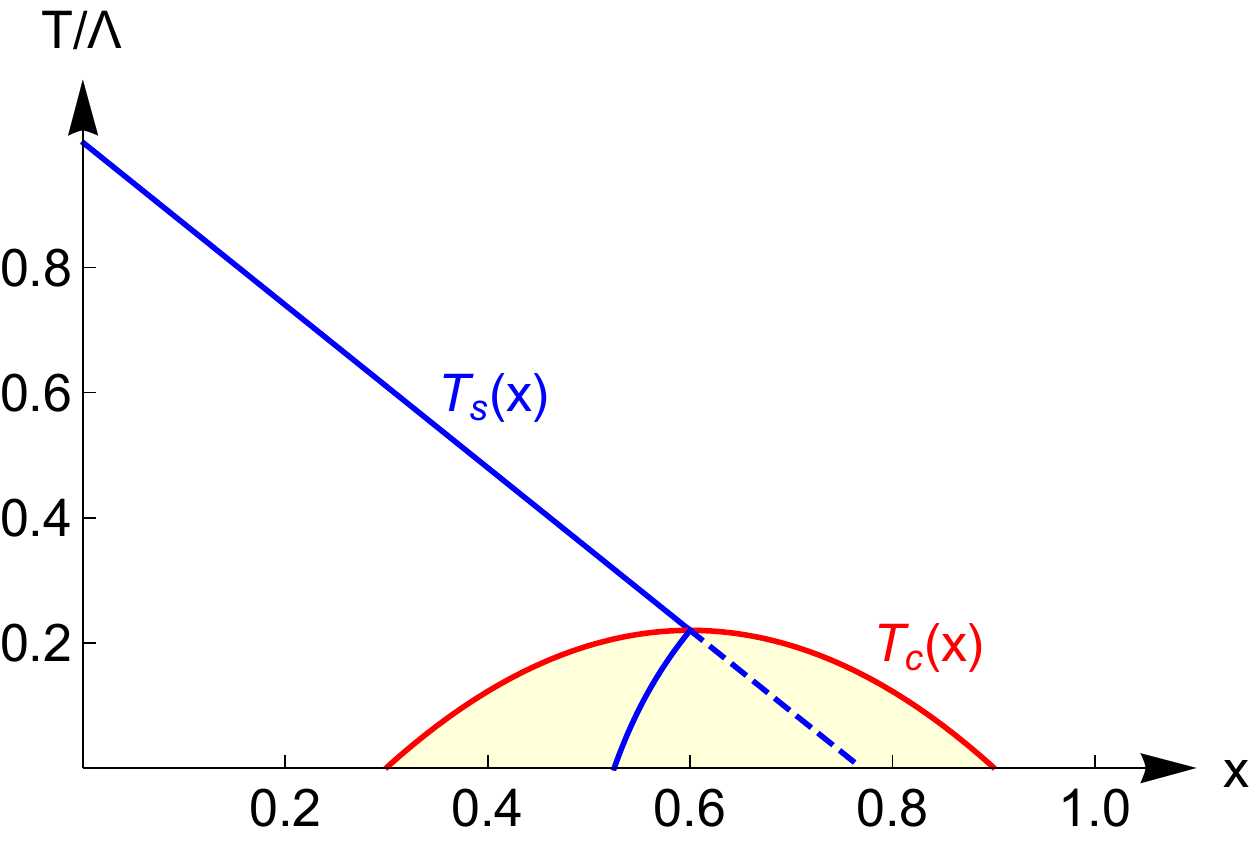}
\caption{
Back-bending of the nematic transition line (solid blue line) $T_s(x)$ in the superconducting phase (shaded light yellow)
due to strong interplay between the two orders. The blue dashed line is the hypothetical nematic transition
if the interplay is ignored.
}
\label{fig3}
\end{center}
\end{figure}
\emph{2. Back-bending of $T_s(x)$ in the superconducting phase}.
As noted above, for $\lambda_0$ greater than a positive threshold the shear modulus $c_s(T)$ hardens for $T \leq T_c$ (red/dark line in
Fig.~(\ref{fig2})). An immediate
consequence of this behavior is the back-bending of the nematic/orthorhombic transition line $T_s(x)$ in the superconducting phase,
as shown in Fig.~(\ref{fig3}). Here $x$ is a hypothetical tuning parameter that, in practice, can be related to doping or pressure.
To illustrate the back-bending we consider the same model of $P(T)$ as above, and we introduce an $x$-dependence to the
temperature scales $T_1(x)/\Lambda = 49.02 + 1.3 x$ and $T_c(x)/\Lambda = 0.22 - 2.44 (x-0.6)^2$, and to the
parameter $t_2(x) = 3 \times 10^{-3} [T_c(x)/\Lambda]^2$. Thus, in this model
$T_c(x)$ has a dome-like structure, and the $T_s(x)$ is linearly decreasing with $x$. The two transition lines meet at
$x=0.6$, and if the interplay is ignored $T_s(x)$ continues the trend (dashed lines Fig.~(\ref{fig3}))
in the superconducting phase. However, once the interplay is taken into account, the hardening of  $c_s(T)$
for $T < T_c$ implies that there cannot be a nematic transition for $x >0.6$ in the superconducting phase.
Moreover, since the hardening increases with lowering $T$,
it necessarily implies that $T_s(x)$ back-bends in the superconducting phase, as reported in electron-doped BaFe$_2$As$_2$~\cite{nandi2010}.

To summarize, we examined the thermodynamic signatures of the interplay between superconducting and
nematic instabilities. In particular, we studied microscopically the properties of the coupling $\lambda$
between the two orders, see Eq.~(\ref{eq:F}). This is related to how the shear elastic constant $c_s(T)$ changes across a
superconducting transition. We explained why in most systems
$\lambda$ (in suitable unit) is small and nearly temperature independent, which leads to $\delta c_s/|c_s^m| \sim 10^{-6}$
as seen in most classes of superconductors.
The situation is different if, due to an imminent nematic instability,
the nematic correlation length $\xi \gg l$, where $l$ is the interatomic distance.
In this case $\lambda \propto (\xi(T)/l)^4$ has strong
$T$-dependence and it can be boosted by several orders of magnitude.
This leads to large  $\delta c_s/|c_s^m| \sim 10^{-1}$, as seen experimentally in the Fe-based and A15
superconductors. If the bare coupling $\lambda_0$ is
above a positive threshold, it leads to hardening of $c_s(T)$ for $T \leq T_c$, and to the back-bending of the
nematic transition line in the superconducting phase, as seen in doped BaFe$_2$As$_2$. Finally, we predict that the
A15 systems have large electronic nematic correlation
which can be revealed using electronic Raman response and elastoresistivity techniques.

\acknowledgments
We are thankful to L. Bascones, G. Blumberg, A. Chubukov, Y. Gallais, F. Hardy, P. Hirschfeld, D. Maslov,
P. Massat, C. Meingast, A. J. Millis, J. Schmalian, K. Sengupta for insightful discussions.
I.~P. acknowledges financial support from ANR grant (ANR-15-CE30-0025).
P.~K. and B.~M.~A. acknowledge support from the Independent Research Fund Denmark, grant
numbers DFF-6108-00096 and DFF-8021-00047B.



\begin{thebibliography}{99}

\bibitem{review-feas} For reviews see, e.g., M. Norman, Physics {\bf 1}, 21 (2008);
D. C. Johnston, Adv. Phys. {\bf 59}, 803 (2010);
G. R. Stewart, Rev. Mod. Phys. {\bf 83}, 1589 (2011);
P. J. Hirschfeld, M. M. Korshunov, and I. I. Mazin, Rep. Prog. Phys. {\bf 74}, 124508 (2011);
A. V. Chubukov, Annu. Rev. of Condens. Matter Phys. {\bf 3}, 57 (2012).

\bibitem{kuo-fisher-2016}
H.-H. Kuo, J.-H. Chu, J. C. Palmstrom, S. A. Kivelson, and I. R. Fisher,
Science {\bf 352}, 958 (2016).

\bibitem{chowdhury-2011}
D. Chowdhury, E. Berg, and S. Sachdev,
Phys. Rev. B {\bf 84}, 205113 (2011);
E.-G. Moon and S. Sachdev,
Phys. Rev. B {\bf 85}, 184511 (2012).

\bibitem{Livanas}
G. Livanas, A. Aperis, P. Kotetes, and G. Varelogiannis,
Phys. Rev. B {\bf 91}, 104502 (2015).

\bibitem{fernandes-2013}
R. M. Fernandes and A. J. Millis,
Phys. Rev. Lett. {\bf 111}, 127001 (2013).

\bibitem{glasbrenner-2015}
J. K. Glasbrenner, I. I. Mazin, H. O. Jeschke, P. J. Hirschfeld, and R. Valent\'{i},
Nat. Phys. {\bf 11}, 953 (2015).

\bibitem{mishra-2016}
V. Mishra and P. J. Hirschfeld,
New J. Phys. {\bf 18}, 103001(2016).

\bibitem{gallais-2016}
Y. Gallais, I. Paul, L. Chauvi\`{e}re, and J. Schmalian,
Phys. Rev. Lett. {\bf 116}, 017001 (2016).

\bibitem{sprau-2017}
P. O. Sprau, A. Kostin, A. Kreisel, A. E. B\"{o}hmer, V. Taufour, P. C. Canfield, S. Mukherjee,
P. J. Hirschfeld, B. M. Andersen, and J. C. S\'{e}amus Davis,
Science {\bf 357}, 75 (2017).

\bibitem{andersen-2017}
D. D. Scherer, A. C. Jacko, C. Friedrich, E. \c{S}a\c{s}io\u{g}lu, Stefan Bl\"{u}gel, R. Valent\'{i}, and B. M. Andersen,
Phys. Rev. B {\bf 95}, 094504 (2017).

\bibitem{classen-2017}
L. Classen, R.-Q. Xing, M. Khodas, A. V. Chubukov,
 Phys. Rev. Lett. {\bf 118}, 037001 (2017).

\bibitem{benfatto-2018}
L. Benfatto, B. Valenzuela, and L. Fanfarillo,
npj Quantum Materials {\bf 3}, 56 (2018).

\bibitem{ando-2002}
Y. Ando, K. Segawa, S. Komiya, and A. N. Lavrov,
Phys. Rev. Lett. {\bf 88}, 137005 (2002).

\bibitem{howald-2003}
C. Howald, H. Eisaki, N. Kaneko, M. Greven, and A. Kapitulnik,
Phys. Rev. B {\bf 67}, 014533 (2003).

\bibitem{hinkov-2008}
V. Hinkov, D. Haug, B. Fauqu\'{e}, P. Bourges, Y. Sidis, A. Ivanov,
C. Bernhard, C. T. Lin, and B. Keimer,
Science {\bf 319}, 597 (2008).

\bibitem{daou-2010}
R. Daou, J. Chang, David LeBoeuf, Olivier Cyr-Choini\`{e}re, Francis Lalibert\'{e}, Nicolas Doiron-Leyraud,
B. J. Ramshaw, R. Liang, D. A. Bonn, W. N. Hardy, and Louis Taillefer,
Nature {\bf 463}, 519 (2010).

\bibitem{sato-2017}
Y. Sato, S. Kasahara, H. Murayama, Y. Kasahara, E.-G. Moon, T. Nishizaki, T. Loew, J. Porras,
B. Keimer, T. Shibauchi and Y. Matsuda,
Nat. Phys. {\bf 13}, 1074 (2017).

\bibitem{auvray-gallais-2019}
N. Auvray, S. Benhabib, M. Cazayous, R. D. Zhong, J. Schneeloch, G. D. Gu, A. Forget, D. Colson, I. Paul,
A. Sacuto, and Y. Gallais,
arXiv:1902.03508.

\bibitem{kivelson-2003}
S. A. Kivelson, I. P. Bindloss, E. Fradkin, V. Oganesyan, J. M. Tranquada, A. Kapitulnik, and C. Howald,
Rev. Mod. Phys. {\bf 75}, 1201 (2003).

\bibitem{vojta-2009}
M. Vojta,
Adv. Phys. {\bf 58}, 699 (2009).

\bibitem{wang-2013}
J. Wang and G.-Z. Liu,
New J. Phys. {\bf 15}, 073039 (2013).

\bibitem{yamase-2013}
H. Yamase and R. Zeyher,
Phys. Rev. B {\bf 88}, 180502(R) (2013).

\bibitem{maier-2014}
T. A. Maier, and D. J. Scalapino,
Phys. Rev. B {\bf 90}, 174510 (2014).

\bibitem{metlitski-2015}
M. A. Metlitski, D. F. Mross, S. Sachdev, T. Senthil,
Phys. Rev. B {\bf 91}, 115111 (2015).

\bibitem{lederer-2015}
S. Lederer, Y. Schattner, E. Berg, and S. A. Kivelson,
Phys. Rev. Lett. {\bf 114}, 097001 (2015).

\bibitem{labat-2017}
D. Labat and I. Paul,
Phys. Rev. B {\bf 96}, 195146 (2017).

\bibitem{klein-2018}
A. Klein, Y.-M. Wu, A. Chubukov,
arXiv:1812.00521

\bibitem{lederer-2019}
S. Lederer, E. Berg, and E.-A. Kim,
arXiv:1908.03224.

\bibitem{fernandes-2010}
R. M. Fernandes, L. H. VanBebber, S. Bhattacharya, P. Chandra, V. Keppens, D. Mandrus, M. A. McGuire,
B. C. Sales, A. S. Sefat, and J. Schmalian,
Phys. Rev. Lett. {\bf 105}, 157003 (2010).



\bibitem{fernandes-2014}
R. M. Fernandes, A. V. Chubukov, and J. Schmalian,
Nat. Phys. {\bf 10}, 97 (2014).

\bibitem{gallais-review-2016}
Y. Gallais and I. Paul,
C. R. Phys. {\bf 17}, 113 (2016).

\bibitem{kushnirenko-2018}
Y. S. Kushnirenko, D. V. Evtushinsky, T. K. Kim, I.V. Morozov, L. Harnagea, S. Wurmehl,
S. Aswartham, A.V. Chubukov, and S. V. Borisenko,
arXiv:1810.04446.

\bibitem{olsen}
J. L. Olsen, Nature {\bf 175}, 37 (1955).

\bibitem{alers61}
G. A. Alers and D. L. Waldorf,
Phys. Rev. Lett. {\bf 6}, 677 (1961).

\bibitem{nohara}
M. Nohara, T. Suzuki, Y. Maeno, T. Fujita, I. Tanaka and H. Kojima,
Phys. Rev. B {\bf 52}, 570 (1995).

\bibitem{bruls}
G. Bruls, D. Weber, B. Wolf, P. Thalmeier, and B. L\"{u}thi,
Phys. Rev. Lett. {\bf 65}, 2294 (1990).

\bibitem{thalmeier}
P. Thalmeier, B. Wolf, D. Weber, G. Bruls, B. L\"{u}thi, and A. A. Menovsky,
Physica C {\bf 175}, 61 (1991).

\bibitem{testardi75}
L. R. Testardi, Phys. Rev. B {\bf 12}, 3849 (1975).

\bibitem{millis-rabe}
A. J. Millis and K. M. Rabe, Phys. Rev. B {\bf 38}, 8908 (1988).

\bibitem{yoshizawa}
M. Yoshizawa and S. Simayi,
Mod. Phys. Lett. {\bf 26}, 1230011 (2012).

\bibitem{zvyagina13}
G. A. Zvyagina, T. N. Gaydamak, K. R. Zhekov, I. V. Bilich, V. D. Fil, D. A. Chareev, and A. N. Vasiliev,
EPL {\bf 101}, 56005 (2013).

\bibitem{boehmer-2014}
A. E. B\"{o}hmer, P. Burger, F. Hardy, T. Wolf, P. Schweiss,
R. Fromknecht, M. Reinecker, W. Schranz, C. Meingast,
Phys. Rev. Lett. {\bf 112}, 047001 (2014).

\bibitem{boehmer-2016}
A. E. B\"{o}hmer and C. Meingast,
C. R. Physique {\bf 17}, 90 (2016).


\bibitem{testardi-67}
L. R. Testardi and T. B. Bateman,
Phys. Rev. {\bf 154}, 402 (1967).

\bibitem{rehwald-72}
W. Rehwald, M. Rayl, R. W. Cohen, and G. D. Cody,
Phys. Rev. B {\bf 6}, 363 (1972).

\bibitem{testardi-rmp}
L. R. Testardi, Rev. Mod. Phys. {\bf 47}, 637 (1975).

\bibitem{koyama04}
T. Koyama,
Phys. Rev. B {\bf 70}, 226503 (2004).

\bibitem{agd}
e.g., see A. A. Abrikosov, L. P. Gorkov, and I. E. Dzyaloshinski,
\emph{Methods of Quantum Field Theory in Statistical Physics},
Dover Publications Inc., New York (1975), chapter 2, pp. 77-78.

\bibitem{chu2012}
J.-H. Chu, H.-H. Kuo, J. G. Analytis, I. R. Fisher,
Science {\bf 337}, 710 (2012).

\bibitem{gallais2013}
Y. Gallais, R. M. Fernandes, I. Paul, L. Chauvi\`{e}re, Y.-X. Yang, M.-A. M\'{e}asson,
M. Cazayous, A. Sacuto, D. Colson, and A. Forget,
Phys. Rev. Lett. {\bf 111}, 267001 (2013).

\bibitem{blumberg2016}
V. K. Thorsmølle, M. Khodas, Z. P. Yin, C. Zhang, S. V. Carr, P. Dai, and G. Blumberg,
Phys. Rev. B {\bf 93}, 054515 (2016).

\bibitem{andersen-2019}
D. Steffensen, P. Kotetes, I. Paul, and B. M. Andersen,
Phys. Rev. B {\bf 100}, 064521 (2019).

\bibitem{wang-2016}
L. Wang, F. Hardy, T. Wolf, P. Adelmann, R. Fromknecht, P. Schweiss, and C. Meingast,
Phys. Status Solidi B {\bf 254}, 1600153 (2017).

\bibitem{nandi2010}
S. Nandi, M. G. Kim, A. Kreyssig, R. M. Fernandes, D. K. Pratt, A. Thaler, N. Ni, S. L. Bud'ko, P. C. Canfield,
J. Schmalian, R. J. McQueeney, and A. I. Goldman,
Phys. Rev. Lett. {\bf 104}, 057006 (2010).


\end{thebibliography}
\end{document}